\documentclass[conference]{IEEEtran}
\IEEEoverridecommandlockouts
\usepackage{cite}
\usepackage{amsmath,amssymb,amsfonts}
\usepackage{algorithmic}
\usepackage{graphicx}
\usepackage{textcomp}
\usepackage{xcolor}

\usepackage{cite}
\usepackage{amsmath,amssymb,amsfonts}
\usepackage{algorithmic}
\usepackage{graphicx}
\usepackage{textcomp}
\usepackage{xcolor}
\usepackage{gensymb}
\usepackage{mathtools, nccmath}
\usepackage{mathtools, nccmath}
\usepackage{lipsum}
\usepackage{amssymb}
\usepackage{pifont}

\usepackage{tabularray}
\usepackage{pbox}
\usepackage[flushleft]{threeparttable}
\usepackage{multirow}
\usepackage{graphicx}
\usepackage{subcaption}
\usepackage[colorlinks,citecolor=black,urlcolor = black, linkcolor = black]{hyperref}

\def\BibTeX{{\rm B\kern-.05em{\sc i\kern-.025em b}\kern-.08em
    T\kern-.1667em\lower.7ex\hbox{E}\kern-.125emX}}
\begin{document}

\title{A Persistent Hierarchical Bloom Filter-based Framework for Authentication and Tracking of ICs}

\author{
   \IEEEauthorblockN{
   \IEEEauthorrefmark{1}Fairuz Shadmani Shishir,
   \IEEEauthorrefmark{1}Md Mashfiq Rizvee,
   \IEEEauthorrefmark{1}Tanvir Hossain,   
\IEEEauthorrefmark{1}Tamzidul Hoque,
\IEEEauthorrefmark{2}Domenic Forte, and \\
\IEEEauthorrefmark{1}Sumaiya Shomaji}
   \IEEEauthorblockA{
   \IEEEauthorrefmark{1}Department of Electrical Engineering and Computer Science, University of Kansas  \\
   \IEEEauthorrefmark{2}Department of Electrical Engineering, University of Florida \\
   Email: \IEEEauthorrefmark{1}\{shishir, mashfiq.rizvee, tanvir, hoque, shomaji\}@ku.edu} \IEEEauthorrefmark{2}\{dforte\}@ece.ufl.edu}
\maketitle
\vspace{-0.2in}
\begin{abstract}
Detecting counterfeit integrated circuits (ICs) in unreliable supply chains demands robust tracking and authentication. Physical Unclonable Functions (PUFs) offer unique IC identifiers, but noise undermines their utility. This study introduces the Persistent Hierarchical Bloom Filter (PHBF) framework, ensuring swift and accurate IC authentication with an accuracy rate of 100\% across the supply chain even with noisy PUF-generated signatures.
\end{abstract}

\vspace{-0.14in}
\section{Introduction}
\vspace{-0.10in}
Ensuring authenticity and quality of electronics is crucial, especially in high-security and reliability applications. Physical Unclonable Functions (PUFs) offer unique digital signatures per chip, but environmental factors can disrupt PUF-based authentication since temperature, aging, and voltage variation cause the output of the PUF (i.e., responses) to change over time for a given input (i.e., challenges). Error correcting codes (ECC) introduce overhead and create security risks by opening the door for side-channel attacks to leak the PUF responses \cite{alaql}. Tracking ICs in real-world volumes poses speed, storage, and security issues. Existing blockchain-based solutions some of these \cite{xu} but neglect errors due to PUF noise. Another solution focuses on IC traceability using blockchain but lacks scalability \cite{rekha}.


Bloom Filters (BF) enable time- and storage-efficient set membership matching \cite{bloom}. In the context of IC tracking, a BF can be used to store the challenge-response pairs (CRPs) of all manufactured ICs. When a new IC is produced, its responses (i.e., signature) for a specific set of challenges can be enrolled in the BF. To check if an IC is authentic, the responses obtained from that IC using the same set of challenges are queried to the BF with all authentic ICs. If the responses are not found there, the IC is likely to be a counterfeit. However, standard BFs cannot deal with noisy data. Therefore, any error in the PUF responses may result in an erroneous membership test. Hierarchical BFs (HBFs) \cite{shomaji}, on the other hand, have a minimal false positive rate and can check queries even if they are noisy. On the other hand, HBFs are inadequate in handling temporal membership; that is, queries with time stamps (e.g., location of chip $X$ at a specific time). While Persistent BFs (PBFs) \cite{peng} can be used to enable such queries, they suffer from the same problem as the BFs; as PBFs cannot deal with noisy data. Utilizing the concepts of the noise tolerance in HBFs combined with the temporal membership query ability in PBFs, in this paper, we propose a noise-tolerant framework with temporal query feature, Persistent Hierarchical BF (PHBF) (fig. \ref{fig:PHBF}).

\section{methodology}
\vspace{-0.05in}
\begin{figure}[!h]
    \centering
    \includegraphics[width=0.85\linewidth]{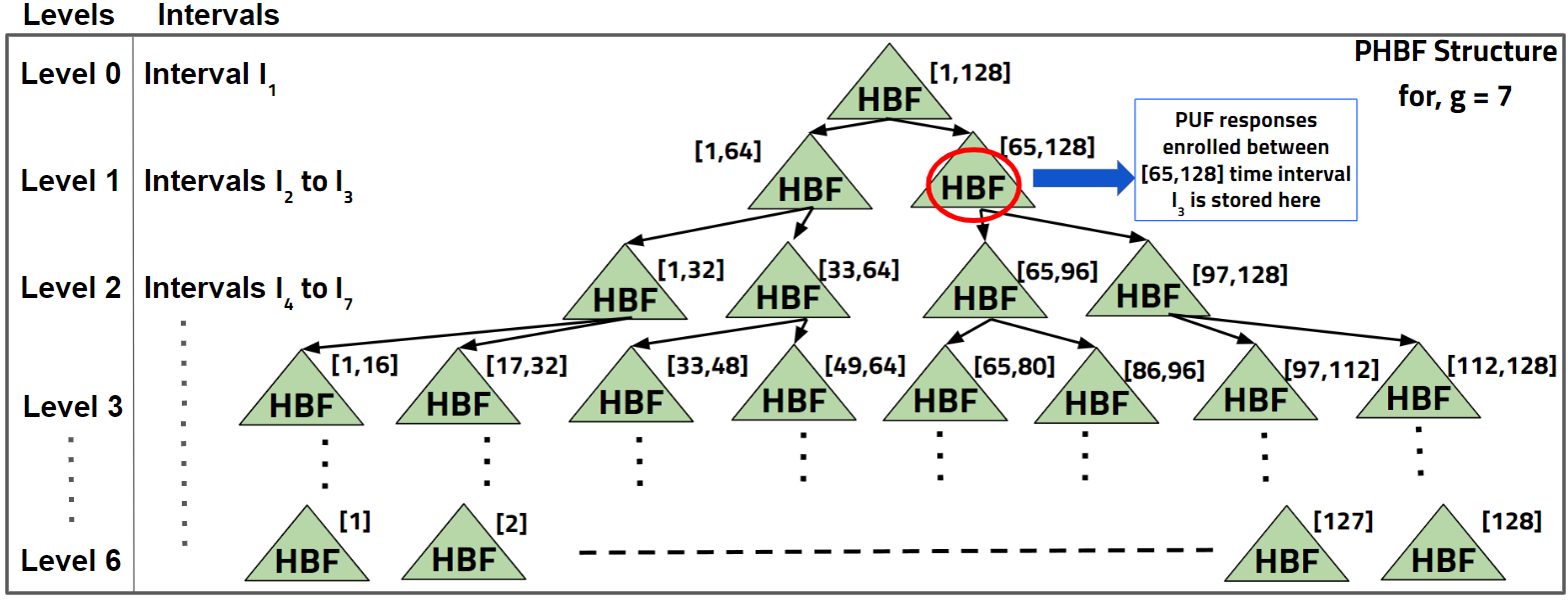}
    \caption{A schematic of the proposed PHBF framework}
    \label{fig:PHBF}
\end{figure}
\textbf{PHBF Construction:} The construction of the proposed PHBF requires the combination of HBF and PBF (fig. \ref{fig:PHBF}). Since HBF is the basis of PHBF, and BF is the basis of HBF, we first initialize the associated parameters of BF and HBF. There are inputs for our proposed framework: number of ICs ($n$), locations ($L$), and days ($T$). For a given $n$, we initialize the no. of BFs ($N$), size of each BF ($m$), no. of hash functions ($k$) from \cite{bloom} and \cite{shomaji}. We also initiate the false positive rate, $FP=$ 10\%. Such parameter initialization serves as the pre-requisite of the PHBF's initialization. Next, utilizing the $T$ input, we decide the no. of levels as well as nodes for the PHBF. Finally, for $L$ no. of locations, we employ $L$ no. of PHBFs. One great feature of the HBF is that it is possible to determine a threshold that helps to achieve a desired authentication performance. As shown in \cite{shomaji}, we tuned the optimum false positive threshold ($FP_{th}$) using:
\vspace{-0.10in}
\textcolor{black}{\begin{equation}
\scriptsize
\begin{aligned}
& FP_{th} = (1-p_{BF}) = \sum_{i=0}^{\left[N_t\right]}\left(\begin{array}{c}
N \\
i
\end{array}\right)\left(1-p_{B F}\right)^{k i}(1-(1-
& \left.\left.p_{B F}\right)^k\right)^{(N-i)}
&
\end{aligned}
\nonumber
\end{equation}}
\vspace{-0.13in}
\begin{table*}[!t]
\scriptsize
    \centering
    \caption{ Proposed query set and their roles for identifying different types of counterfeits.}
        \label{relationship table for counterfeits}
\begin{tabular}{|p{0.05\linewidth}|p{0.15\linewidth}|p{0.25\linewidth}|p{0.42\linewidth}|}
    \hline
        \textbf{No.} & \textbf{Counterfeit Type} & \textbf{Query} & \textbf{Relation Between Query and  Counterfeit Type Determination}   \\ \hline \hline
        1 & Theft & What was ChipX’s last location? &Integrated circuits (ICs) are expected to move through specific stages in the supply chain. Therefore, if the chip visits every location of the presumed trajectory and we find it  enrolled in every PHBF, we can say that the chip is not missing or stolen. \\ \hline
        2 & Cloned or \newline overproduced & Is ChipX enrolled at the OEM’s PHBF? & In this case, if the chip marking matches but the response do not, we can conclude that the chip is cloned or overproduced. To detect this type of counterfeit, performing a query only at the OEM-end PHBF is enough since it is the starting point of the supply chain. \\ \hline
        3 & Remarked & Is ChipX present at OEM’s PHBF but the external marking is different than that? & Remarked ICs could be detected if the chip response matches but the marking does not when a query is performed at the OEM’s end. \\ \hline
        4 & Recycled & Has ChipX been sold to an end user? & In order to keep track of the sold ICs, we keep an HBF at the OEM’s end so that the responses could be recorded whenever a chip is sold to an end user. If our query chip response already exists into the HBF, we can say that the chip is recycled since it was previously purchased by someone. \\ \hline
\end{tabular}
\label{tab: table2}
\vspace{-0.27in}
\end{table*}

\noindent {where $P_{BF}$ is the probability of having a falsely-set-to-1 bit in a BF, $k$ is the total number of hash functions, and $i$ is the number of levels in HBF. Once the parameters associated with BF and HBF are determined, $T$ is utilized to complete the remaining PHBF initialization process. The consideration of $T$ which is a universal set of days means that the total time frame that we are going to cover is $T$ days.  We restrict the time span using time granularity, $g$ where $g \textless log(T)$ and $2 ^ {g} \le T$ to control the number of leaf nodes of intervals in the tree. Binary decomposition of time-span $[1,T]$ consists $Level = \lceil(log(\frac{T}{g}))\rceil + 1$ levels indexed by $l = 0,1,2,\dots, Level-1$ where $Level$ 0 is the root node and $Level-1$ contains the leaf level with $[\frac{T}{g}]$ intervals each of length g.  Level $l$ for each $l \in [0,Level-1]$ contains $2^l$ intervals of length $T/2^{l}$ each. As per the formula mentioned, level 0 or \textcolor{black}{the root node contains} \textcolor{black}{longest time interval $\{[1,T]\}$ $\in$ $ T$. Subsequently,} level 1 contains $\{[1,\frac{T}{2}],[\frac{T}{2}+1,T]\}$, and level $l$ contains $\{[1,\frac{T}{2^l}],[\frac{T}{2^l} +1, 2\frac{T}{2^l}],\dots,[T-\frac{T}{2^l} +1,T]\}$ time intervals. Thus, the total number of temporal interval is $u = (2({T}/{g}) -1)$. A binary tree is formed using temporal intervals. The intervals are indexed in a level-ordered fashion which leads to an interval set $I = \{I_{1},I_{2},\dots,I_{u}\}$. To summarize, PHBF is organized into $Level$s and consists of HBFs $\{hbf_{1}, hbf_{2},\dots,hbf_{u}\}$ such that level $l$, for $l \in [0,Level-1]$, consists of HBFs $\{hbf_{2^{l}},\dots,hbf_{2^{l+1}-1}\}$. Each of the HBFs is constructed with $l_{0}$ length of BFs with m bits. Now, if we consider the number of enrolled responses $n$, we achieve the optimal number of hash functions using:}
\vspace{-0.05in}
\small
\begin{equation}
  k = \left(\frac{m}{n}\right)ln2
\vspace{-0.05in}
\end{equation}
\normalsize
In order to achieve irreversibility, a cryptographic hash function, SHA256 was used combined with a non-cryptographic modulus function, FNV to truncate further\cite{fnv}.

\textbf{Enrollment and Query Processing in PHBF:} {In PHBF, the insertion and query processing is similar to segment tree operations. When enrolling an element ``x'' at a timestamp ``t'', we use Depth First Search \cite{tarjan} over the binary decomposition tree to find a path from the root to the leaf containing ``t" and insert ``x" to every HBF along the path. In case of a query $Q(x,[s,e])$ where $[s,e]$ is a time range and $[s,e] \subset I$, we use DFS to find the binary decomposition of the range $[s,e]$ and query each HBF of the intervals. The query is successful only if at least one HBF returns TRUE for the membership testing of ``x". Formally, we first define the canonical cover of range $Cover([s,e])= \{I_{a_{1}},\dots,I_{a_{i}}\}$, where $a_{i}\in[1,u]$ are the index values of the intervals that are contained in the cover. To illustrate, considering $Cover([s,e])$ covers the query interval entirely, we answer $Q(x,[s,e])$; where $q(x,hbf_{i})$ means a membership test of ``x” at $hbf_{i}$:}
\vspace{-0.02in}
\small
\begin{equation}
     Q\left(x,[s,e]\right) = 
     Q\left(x,hbf_{a_{1}}\right) \lor Q\left(x,hbf_{a_{2}}\right) \lor 
    \dots Q\left(x,hbf_{a_{i}}\right) 
    \end{equation}
\normalsize
\noindent\textcolor{black}{For instance, if we want to find $Cover([17,48])$, illustrated in fig. \ref{fig:PHBF}, the result is $[I_9,I_{10}]$. Since, $I_9=[17,32]$, and $I_{10}=[33,48]$ we will get our membership test answer of ``x”s membership test in PHBF using:}
\small
\begin{equation}
Q\left(x,[17,48]\right) = Q\left(x,hbf_{9}\right) \lor Q\left(x,hbf_{10}\right)
\end{equation}
\normalsize
\noindent{If  at least one of the HBFs returns TRUE, the data structure considers ``x" to be found.} 

\textbf{Role of PHBF to Detect Counterfeit ICs:}
Using our data structure, it is possible to detect at least 4 types of IC counterfeits: Theft, Cloned/Overproduced, Remarked, and Recycled. We propose a set of queries that helps us determine if a particular IC is a counterfeit or not. The proposed set of queries are shown in Tab. \ref{tab: table2}.

\section{Results}
\vspace{-0.05in}


\begin{figure}[!h]
    \centering
    \includegraphics[width=0.92\linewidth]{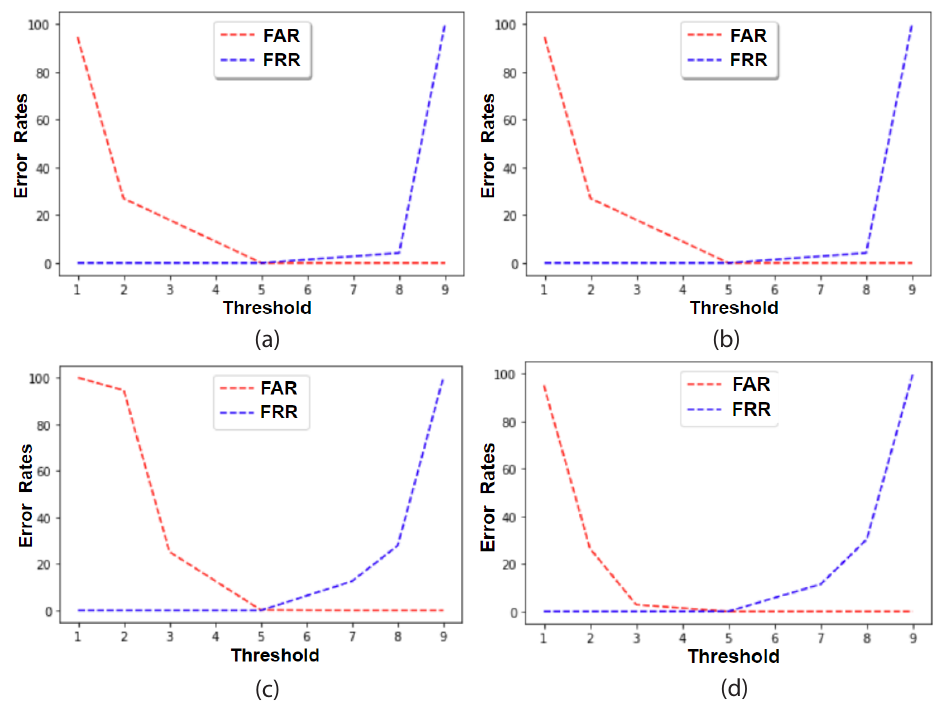}
    \caption{Performance evaluation of PHBF using ROC curve with four query datasets: (a) dataset containing real noise (75\textdegree C), (b) dataset containing real noise (100\textdegree C), (c) dataset containing synthetic noise (clustered), and (d) dataset containing synthetic noise (uniform).}
    \label{fig:HBF_results}
\end{figure}

\subsection{Evaluation Metric}

Using four of the prepared query sets, we performed multiple PUF response authentication experiments on the HBF and the PHBF. Results are represented using the Receiver Operating Characteristics curve (fig. \ref{fig:HBF_results}). In all the cases, our framework achieved a 100\% authentication rate when we set our threshold, $th=5$.

\section{Conclusion}
\vspace{-0.05in}
We have presented a novel framework PHBF which is a fast and noise-tolerant framework that offers robust authentication and tracking of ICs. Future work includes the inclusion of deletion property in the proposed PHBF framework and thorough scalability as well as security analysis.
\vspace{-0.115in}
\bibliographystyle{plain}
\bibliography{bibfile}

\end{document}